\RequirePackage{ifpdf}
\ifpdf 
\documentclass[pdftex]{sigma}
\else
\documentclass{sigma}
\fi

\usepackage{graphicx}

\usepackage{float}
\usepackage{savesym}
\usepackage{amsfonts,amssymb}
\usepackage{amsmath}
\usepackage{color}
\newtheorem{prop}{Proposition}

\usepackage[title]{appendix}

\numberwithin{equation}{section}

\begin{document}

\renewcommand{\PaperNumber}{***}


\ShortArticleName{}

\ArticleName{Smooth multisoliton solutions of the Geng-Xue equation}

\Author{Nianhua LI~$^{\rm a}$ and Q.P. LIU~$^{\rm b}$}

\AuthorNameForHeading{N.~Li and Q. P.~Liu}

\Address{$^{\rm a)}$~School of Mathematical Sciences, Huaqiao University,
Quanzhou, 362021, P R China \\
Faculty of Mathematics, National Research University Higher School of Economics, 119048, Moscow, Russia} 
\EmailD{\href{mailto:email@address}{linianh@hqu.edu.cn}} 

\Address{$^{\rm b)}$~ Department of Mathematics, China University of Mining and Technology, Beijing,
100083, P R China }
\EmailD{\href{mailto:email@address}{qpl@cumtb.edu.cn}} 



\Abstract{We present a reciprocal transformation which links the Geng-Xue equation to a particular reduction of the first negative flow of the Boussinesq hierarchy. We discuss two reductions of the reciprocal transformation for the Degasperis-Procesi and Novikov equations, respectively. With the aid of the Darboux transformation and the reciprocal transformation,
we obtain a compact parametric representation for the smooth soliton solutions such as multi-kink solutions of the Geng-Xue equation.}


\Classification{35Q51; 35C08; 37K10} 

\section{Introduction}
The Degasperis-Procesi (DP) equation
\begin{equation}\label{dp}
m_t+3u_xm+um_x=0,\quad m=u-u_{xx}
\end{equation} was derived by applying the method of asymptotic integrability to  a many-parameter family of third order dispersive PDEs \cite{Procesi}.
It may be viewed as an approximate model describing shallow water wave propagation in the small amplitude and long wavelength regime  waves \cite{johnson,dgh,Lannes,Iva}.
The DP equation is a completely integrable equation. It admits a Lax pair, bi-Hamiltonian structure, and is reciprocally related to a negative flow in the Kaup-Kupershmidt hierarchy \cite{Degasperis}.
 Moreover, the DP equation has been studied by the inverse scattering method \cite{Constantin,Boutet}, and was shown to possess various periodic-wave solutions and travelling-wave solutions \cite{Vak,Lenells}.
Its smooth multi-soliton solutions were constructed by means of the $\tau$ function approach,   Riemann-Hilbert method,   dressing method
and   Darboux transformation (DT) \cite{Boutet,Matsuno1,Ivanov,Li2}.
In particular, the DP equation is an equation of Camassa-Holm (CH) type, which has an unusual feature of admitting  non-analytic solutions called peakons \cite{Szmigielski,Szmigielski2}. There exists an interesting connection between the DP peakon
lattice and the finite C-Toda lattice \cite{Chang}.

By using the approach of perturbative symmetry to classifying  integrable equations of CH form, another CH type equation with cubic nonlinearity
\begin{equation}\label{novikov}
m_{t}+u^{2}m_{x}+3uu_{x}m=0,\quad m=u-u_{xx}
\end{equation}
 was discovered by Vladimir Novikov \cite{Novikov}. Soon afterwards, Hone and Wang \cite{Hone} confirmed  its integrability by presenting a Lax representation, infinitely many conserved
quantities as well as  a bi-Hamiltonian structure. They also related this equation to a negative flow of  the Sawada-Kotera hierarchy via a reciprocal transformation. Smooth
multi-soliton solutions of the Novikov equation have been presented via several approaches such as the Hirota bilinear method, Riemann-Hilbert method and  DT \cite{Matsuno2,Shepelsky,Wu}. Furthermore, multipeakons of the Novikov equation may be computed by inverse spectral method \cite{Lundmark}. Dynamical system of the multipeakons of the Novikov equation is a Hamiltonian system, which is connected to the finite Toda lattice of BKP type \cite{Hone,Zhao}.

Subsequently, Geng and Xue \cite{Geng} proposed a two-component generalization of the Novikov equation and the DP equation
\begin{equation}\label{novikov2}\left\{
\begin{array}{rr}
 m_{t}+3u_{x}vm+uvm_{x} =0,& \\
 n_{t}+3v_{x}un+uvn_{x} =0,&  \\
 m=u-u_{xx},\quad   n=v-v_{xx}.& 
\end{array}\right.\end{equation}
Indeed, for $u=1$ and $u=v$, the Geng-Xue equation (\ref{novikov2}) reduces to the DP equation and the Novikov equation,  respectively. This equation is  a completely integrable system  with a Lax pair and bi-Hamiltonian structure \cite{Geng,Liu}. It is mentioned that the homogeneous and local properties of the Hamiltonian functionals were discussed \cite{Lihong}. Also, the Geng-Xue equation is related to a negative flow in a modified Boussinesq hierarchy by a reciprocal transformation \cite{Li} and the behaviour of the  bi-Hamiltonian structures under the transformation was studied \cite{Lihong2}.
 Moreover, the Geng-Xue equation was shown to admit multi-peakon solutions
\cite{Lund2,Szmigiel,Shuai} and its Cauchy problem was considered \cite{ZRLiu,hm}. However, to the best of our knowledge, smooth solutions such as multi-soliton solutions of the Geng-Xue equation have not been constructed.

The purpose of this paper is to propose a method for building soliton solutions of the Geng-Xue equation. To this end, we find it is convenient to relate the Geng-Xue equation  to  a particular reduction of the first negative flow in the Boussinesq hierarchy via a reciprocal transformation.
Different from the works of \cite{Li} and \cite{Lihong2},
this reciprocal transformation may be reduced to that of the DP equation and the Novikov equation. Furthermore, by
combining the reciprocal transformation with DT of the negative flow in the Boussinesq hierarchy,
we are able to obtain a parametric representation for the multi-kink solutions of the Geng-Xue equation.

The paper is arranged as follows. In  section 2, we introduce a reciprocal transformation  and relate the Geng-Xue equation to a particular reduction of the first negative flow of the Boussinesq hierarchy.
The reductions of the reciprocal transformation to the DP equation and the Novikov equation will also be discussed. In section 3, with the aid of the reciprocal transformation and Darboux transformation, we construct the smooth multisoliton or  multi-kink solutions of the Geng-Xue equation. Interestingly  the solutions will be represented in terms of Wronksians, and the simplest nontrivial cases will be given explicitly.

\noindent
\section{A reciprocal transformation and negative flow of the Boussinesq hierarchy}
In this section, we present a proper reciprocal transformation and establish a link between the Geng-Xue equation and a special negative flow of the Boussinesq hierarchy. Also, we consider the possible reductions of the reciprocal transformation  and show that the reciprocal transformations for both the DP equation and Novikov equation are recovered.

\subsection{A reciprocal transformation of the Geng-Xue equation}
The Geng-Xue equation (\ref{novikov2}) has a Lax representation \cite{Geng}, namely it is the compatibility condition of
\begin{equation}\label{lax-1}
 \Phi_{x}=M\Phi, \quad \;
 \Phi_{t}=N\Phi,
\end{equation}
where $\Phi=(\varphi_1, \varphi_2, \varphi_3)^T$ and
\[
\hspace{-1.0cm}M=\left(
    \begin{array}{ccc}
      0 & m\lambda & 1 \\
      0 & 0 & n\lambda \\
      1 & 0 & 0 \\
    \end{array}
  \right), \quad \; N=\left(
    \begin{array}{ccc}
      -u_{x}v & \frac{u_{x}}{\lambda}-uvm\lambda & u_{x}v_{x} \\
      \frac{v}{\lambda}& u_{x}v-uv_{x}-\frac{1}{\lambda^{2}} & -uvn\lambda-\frac{v_{x}}{\lambda} \\
      -uv & \frac{u}{\lambda}& uv_{x} \\
    \end{array}
  \right).
\]
Use of the above linear spectral problem (\ref{lax-1}) and a standard algorithm lead to infinitely
many conservation laws for the Geng-Xue equation.
One of them is
\[
h_t=(uvh)_x,\;\quad h=(mn)^{\frac{1}{3}},
\]
which allows us to introduce new independent variables $y$ and $\tau$ via the following reciprocal transformation
\begin{equation}\label{rec1}
dy=hdx-uvhdt, \quad \; d\tau=dt.
\end{equation} Setting $w=(\frac{m}{n})^{\frac{1}{3}}$ and eliminating $\varphi_1,\varphi_2$, we find  that the linear system (\ref{lax-1})
may be rewritten in terms of $\varphi\equiv\varphi_3$ as
\begin{equation}\label{scalargeng}
\left\{\begin{array}{rr}\left[(wh)^{-\frac{3}{2}}(h(h\varphi_{y})_y-\varphi)\right]_y-\lambda^2h^{\frac{1}{2}}w^{-\frac{3}{2}}\varphi=0,&\\[5pt]
\varphi_{\tau}-\lambda^{-2}u(wh)^{-\frac{3}{2}}[h(h\varphi_{y})_y-\varphi]-uv_yh\varphi=0.&
\end{array}\right.\end{equation}
To bring (\ref{scalargeng}) into a familiar form,  we introduce a gauge  transformation
\begin{equation}\label{gu2}
\varphi=h^{-\frac{1}{2}}w^{\frac{1}{2}}\phi,
\end{equation}
and have
\begin{equation}\label{lax-2}
\left\{
\begin{array}{l}
\phi_{yyy}+\xi\phi_y+\eta\phi=\lambda^2\phi, \\[5pt]
\phi_{\tau}-\frac{p}{\lambda^2}\left[\phi_{yy}+q\phi_y+(\xi-q_y+q^2)\phi\right]=0,
 \end{array}\right.
\end{equation}
where
\begin{equation}\label{p-q}
p=uh^{\frac{1}{2}}w^{-\frac{3}{2}}, \quad q=\frac{w_y}{w}
\end{equation}
 and
\begin{eqnarray*}
\xi=\frac{3w_{yy}}{2w}-\frac{9w_y^2}{4w^2}-\frac{h_{yy}}{2h}+\frac{h_y^2-4}{4h^2},\;
\eta=\xi_y-\frac{w_{yyy}+\xi w_y}{w}+\frac{6w_yw_{yy}}{w^2}-\frac{6w_y^3}{w^3}.
\end{eqnarray*}

It is noted that the first equation of (\ref{lax-2}) is the linear spectral problem of the Boussinesq hierarchy. Now, the compatibility condition of the two equations of (\ref{lax-2}) yields the associated Geng-Xue equation, which reads
\begin{eqnarray}\label{asgx}\left\{
\begin{array}{l}
\xi_{\tau}=-3p_y,\quad \quad \quad \quad \quad ~\;  \ s_{1y}=0, \\
\eta_{\tau}=-3p_{yy}-3(pq)_y,\quad \quad  s_2=0,
 \end{array}\right.
\end{eqnarray}
where
\begin{equation*}
s_1= p_{yy}+3p_yq+3pq^2+\xi p+1, \quad
  s_2=\eta-\xi_y+q_{yy}-3qq_y+\xi q+q^3.
\end{equation*}
It is mentioned that the Geng-Xue equation under the transformation (\ref{rec1}) implies the associated Geng-Xue equation (\ref{asgx}). As system (\ref{asgx}) possesses a Lax pair,  one may construct its infinitely many conserved quantities. Furthermore, its WTC Painlev\'e property may be verified directly.

As mentioned, the spatial part of the spectral problem (\ref{lax-2}) is the one for the Boussinesq hierarchy, thus the associated Geng-Xue equation (\ref{asgx}) should have connection with a particular flow of this hierarchy.
To see it, let us consider a more general spectral problem
\begin{equation}\label{nebou1}
\left\{\begin{array}{l}
\phi_{yyy}+\xi\phi_y+\eta\phi=\lambda^2\phi, \\[5pt]
\phi_{\tau}-\frac{1}{\lambda^2}\left[a\phi_{yy}+b\phi_y+\frac{1}{3}(2a\xi-3b_y-a_{yy})\phi\right]=0.
\end{array}\right.\end{equation}
Its compatibility condition yields
\begin{equation}\label{nbou}
\left(\begin{array}{c} \xi \\
    \eta
      \end{array}\right)_{\tau}=J_1\left(
                           \begin{array}{c}
                             a \\
                             b \end{array}
                         \right),~~ \left(
                                         \begin{array}{c}
                                           z_1 \\
                                           z_2 \end{array}
                                       \right)\equiv J_2\left(
                           \begin{array}{c}
                             a \\
                             b   \end{array}
                         \right)=0,
 \end{equation}
where
\begin{align*}
J_1&=\left(
                                                                             \begin{array}{cc}
                                                                               -3\partial & 0 \\
                                                                               -3\partial^2 & -3\partial  \end{array}
                                                                           \right
),\\
 J_2&=\left(
              \begin{array}{cc}
               \partial^4+\xi\partial^2-\partial \xi_y+3\eta\partial+2\eta_y & 2\partial^3+\xi\partial+\partial \xi\\
                \frac{1}{3}(\partial^5+\xi\partial^3-2\partial^3\xi-2\xi\partial \xi)+3\partial \eta\partial+\eta_{yy} & \partial^4+\xi\partial^2+3\eta\partial+\eta_y
              \end{array}
            \right).
\end{align*}It is not difficult to check that $J_2J_1^{-1}$ is the well-known recursion operator of the Boussinesq hierarchy, and hence the system (\ref{nbou}) is just the first negative flow in the  Boussinesq hierarchy.
Now, setting $a=p, b=pq$, we have the following relation
\begin{equation*}
\left(
  \begin{array}{c}
    z_1 \\
    z_2
  \end{array}
\right)=\left(
            \begin{array}{cc}
              \partial-q & 2p\partial+3p_y \\
              \frac{1}{3}\partial^2+(q_y-\frac{2}{3}\xi-q^2) & p\partial^2+(3p_y+pq)\partial+3p_{yy}+3(pq)_y
            \end{array}
          \right)\left(
                   \begin{array}{c}
                     s_{1y} \\
                     s_2
                   \end{array}
                 \right).
\end{equation*}
Thus, the associated Geng-Xue equation (\ref{asgx}) indeed  is  a particular reduction of the negative Boussinesq equation (\ref{nbou}).

\subsection{Reductions of the reciprocal transformation}
Recall that the Geng-Xue equation may be reduced to the DP equation and the Novikov
equation as $u= 1$ and $u=v$, respectively. So it is interesting to consider the reciprocal transformation (\ref{rec1}) under these reductions.

\bigskip

 \noindent {\bf Case 1:} $u=1$
 \bigskip

 \noindent In this case, the reciprocal transformation (\ref{rec1}) becomes
\begin{equation}
dy=hdx-vhdt, ~~ d\tau=dt,
\end{equation}where $h=n^{\frac{1}{3}}$. A direct calculation shows that the associated equation (\ref{asgx}) and  the spectral problem (\ref{lax-2}) reduce to
\begin{equation}\label{aDP}
\xi_{\tau}=-6hh_y,~~ 2h_{yyy}+h\xi_y+2\xi h_y=0,
\end{equation}
and
\begin{equation}
\left\{\begin{array}{l}
\phi_{yyy}+\xi\phi_y+\frac{1}{2}\xi_y\phi=\lambda^2\phi,\\[5pt]
\phi_{\tau}=\frac{1}{\lambda^2}\left[h^2\phi_{yy}-hh_y\phi_y+(h_y^2-hh_{yy}-1)\phi\right],\end{array}\right.
\end{equation}
respectively.
 These are just the associated DP equation and its spectral problem \cite{Degasperis,Li2}.

\bigskip
 \noindent{\bf Case 2:} $u=v$
\bigskip

 \noindent
 Now, the reciprocal transformation (\ref{rec1}) turns into
\begin{equation}\label{rec3}
dy=hdx-u^2hdt, ~~ d\tau=dt,
\end{equation}where $h=m^{\frac{2}{3}}$. In this case, the spectral problem (\ref{lax-2}) becomes
\begin{equation}\label{ask}
\phi_{yyy}+\xi\phi_y+\xi_y\phi=\lambda^2\phi,~~  \phi_{\tau}=\frac{p}{\lambda^2}(\phi_{yy}+\xi\phi),
\end{equation}
where
\begin{eqnarray*}
\xi=-\frac{h_{yy}}{2h}+\frac{h_y^2}{4h^2}-\frac{1}{h^2},~~  p=uh^{\frac{1}{2}}.
\end{eqnarray*}
Meanwhile, (\ref{asgx}) yields
\begin{equation}\label{asso-n}
\xi_{\tau}=-3p_y,~~ (p_{yy}+p\xi)_y=0.
\end{equation}
Setting $\phi=\psi_y$, from (\ref{ask}) we have
\begin{equation}\label{asso-nlax}
\psi_{yyy}+\xi\psi_y=\lambda^2\psi,~~
\psi_{\tau}=\frac{1}{\lambda^2}(p\psi_{yy}-p_y\psi_y+p_{yy}\psi+p\xi\psi).
\end{equation}
It is easy to see that by integrating the second equation of (\ref{asso-n}) and subtituting it into (\ref{asso-nlax}), we reach the results appeared in \cite{Hone,Wu}.

%
\section{Multisoliton solutions of the  Geng-Xue equation}
In the previous section, we have related the Geng-Xue equation (\ref{novikov2}) to the associated equation (\ref{asgx}),
which was shown to be a reduction of the first negative flow in the Boussinesq hierarchy. In what follows, we will take a similar approach as done for the CH, modified CH and DP equations \cite{LiYS,Xia,Li2} and explain that  this connection, together with the DT for the associated Geng-Xue equation, allows us to propose an algorithm to build multisoliton solutions of the Geng-Xue equation.

 As a first step, we have the following \par 

%
%
%
\begin{prop} The spectral problem (\ref{lax-2}) is covariant with respect to the following $N$-DT
\begin{align}\label{dt1}
  \phi[N]&=\frac{W(f_1,f_2,...,f_N,\phi)}{W_N},\nonumber\\
  \xi[N]&=\xi+3({\rm ln} W_N)_{yy},\nonumber\\
\eta[N]&=\eta+\left(N\xi +\frac{3}{2}[({\rm ln} W_N)_y]^2+3({\rm ln} W_N)_{yy}-\frac{3F_N}{W_N}\right)_y, \\
  p[N]&=p-({\rm ln} W_N)_{y\tau}, \nonumber\\
  q[N]&=\frac{1}{p[N]}\left(pq+Np_y-({\rm ln} W_N)_y({\rm ln} W_N)_{y\tau}+(\frac{F_N}{W_N})_{\tau}\right),\nonumber
\end{align} where $f_{1}, ..., f_N$ are
 solutions of the spectral problem (\ref{lax-2}) at $\lambda=\lambda_1, ..., \lambda_N$, respectively. $W$ signifies the Wronskian, $W_N=W(f_1,f_2,...,f_N)=|\widehat{N-1}|$ and
\begin{eqnarray*}
F_N=\left\{\begin{array}{rl}
0,  \quad \quad \quad \quad \quad \quad \quad \quad \quad N=1,\\[8pt]
W(f_{1y},f_{2y}),   \quad \quad \quad \quad \quad N=2,\\[8pt]
|\widehat{N-3},N-1,N|,   \quad  \quad N\geq3.
\end{array} \right.
\end{eqnarray*}Here and in the sequel all determinant notations are adopted according to \cite{Freeman}.
\end{prop}

{\it Proof}. For the Lax operator $L=\partial_y^3+\xi\partial_y+\eta$, it is well known that the operator of the 1-DT  is given by $T_1=\partial_y-\frac{f_{1y}}{f_1}$ (see \cite{leble,Matveev}). It is straightforward to check that the transformed variables solve
\begin{equation*}\label{lax2}\left\{
\begin{array}{l}
(\phi[1])_{yyy}+\xi[1](\phi[1])_y+\eta[1]\phi[1]=\lambda^2\phi[1], \\[10pt]
(\phi[1])_{\tau}-\frac{p[1]}{\lambda^2}\Big((\phi[1])_{yy}+q[1](\phi[1])_y+(\xi[1]-(q[1])_y+q[1]^2)\phi[1]\Big)=0,
 \end{array}\right.
\end{equation*}
namely the proposition is valid for $N=1$ case. For the general case, we may assume that the operator of the $N$-DT takes the form $T=\partial_y^N+a_1\partial_y^{N-1}+\cdots+a_N$, where coefficients $a_1,...,a_N$ are functions of $f_1,...,f_N$ and their derivatives with respect to $y$.
Hence, for this iterated DT, we have
\begin{equation}\label{dtop}
\hat{L}T=TL,  \quad    \hat{L}=\partial_y^3+\xi[N]\partial_y+\eta[N].
\end{equation}
Plugging the expressions of $T, L$ into the first equation of (\ref{dtop}) and considering the coefficients of  $\partial^{N+1}$ and $\partial^{N}$, we obtain
\begin{equation}\label{pdt}
\xi[N]=\xi-3a_{1y},\quad   \eta[N]=\eta+N\xi_y+3a_1a_{1y}-3a_{1yy}-3a_{2y}.
\end{equation}
The coefficients of the operator $T$ may be determined from the conditions $Tf_i=0, 1\leq i\leq N$. In other words, we have
\begin{equation}\label{lineq}     
a_Nf_i+a_{N-1}f^{(1)}_{i}+...+a_1f_i^{(N-1)}=-f_i^{(N)}, \; i=1,..., N.
\end{equation}Using Cramer's rule, we obtain from the system (\ref{lineq}) that
\begin{equation}
a_1=-\frac{|\widehat{N-2},N|}{|\widehat{N-1}|},\quad   a_2=\frac{|\widehat{N-3},N-1,N|}{|\widehat{N-1}|}.
\end{equation}Substituting them into (\ref{pdt}), we have
\begin{equation*}
 \xi[N]=\xi+3({\rm ln} W_N)_{yy},\quad \eta[N]=\eta+\left(N\xi +\frac{3}{2}[({\rm ln} W_N)_y]^2+3({\rm ln} W_N)_{yy}-\frac{3F_N}{W_N}\right)_y.
\end{equation*}
In addition, the iterated formulations for $p,q$ may be acquired by substituting $\xi[N],\eta[N]$ into the associated Geng-Xue equation (\ref{asgx}). This completes the proof.
\bigskip

Next, let us start with the trivial solution $u=u_0, v=v_0$ of the Geng-Xue equation, where $u_0, v_0$ are positive constants.
Then the corresponding seed solution in the above DT   reads $\xi=-k^{-2},p=k^2, \eta=q=0$, where $k=(u_0v_0)^{\frac{1}{3}}$.
To calculate the solutions $f_j$ of the system (\ref{lax-2}) at $\lambda_j$ $(j=1,..., N)$, it is convenient to assume $\alpha_j,\beta_j, -\alpha_j-\beta_j$ being three distinct roots of the cubic equation
 $\omega^3-k^{-2}\omega=\lambda_j^2$.
 Now, without loss of generality, from  (\ref{lax-2}) we have 
\begin{equation}\label{fj}
f_{j}=e^{\vartheta_j}+\delta_je^{\sigma_j}, \quad    1\leq j\leq N,
\end{equation} where
\begin{equation*}
\vartheta_j=\alpha_j y+\frac{k^2}{\alpha_j}\tau+c_{1j}, \quad   \sigma_j=\beta_jy+\frac{k^2}{\beta_j}\tau+c_{2j}.
\end{equation*}Herein $c_{1j},c_{2j},\delta_j$ are arbitrary constants.
To build  real solutions of the Geng-Xue equation, let us assume $\alpha_j-\beta_j=p_j>0$ and rewrite (\ref{fj}) as
\begin{equation*}
f_{j}=e^{\frac{\vartheta_j+\sigma_j}{2}}(e^{\theta_j}+\delta_je^{-\theta_j}),\quad
\theta_j=\frac{1}{2}\left(p_jy+\frac{3k^2p_j}{p_j^2-k^{-2}}\tau+c_{1j}-c_{2j}\right).
\end{equation*}

It is noted that  $1$, $e^{y/k}$ and $e^{-y/k}$ constitute a fundamental set of  solutions of the first equation of (\ref{lax-2}) at $\lambda=0$ with the seed $\xi=-k^{-2},p=k^2, \eta=q=0$. With the help of the $f_j$'s  given by (\ref{fj}), it follows immediately from the Proposition 1 that for the $N$-th iterated spectral problem
\[
(\phi[N])_{yyy}+\xi[N](\Phi[N])_y+\eta[N]\Phi[N]=0
\]
at $\lambda=0$ we may calculate its solutions, and in particular we have
\begin{equation}\label{dt2}
  \left\{\begin{array}{l}
\phi[N]=\nu_1\phi^1[N]+\nu_2\phi^2[N]+\nu_3\phi^3[N],\\[3pt]
   p[N]=k_1^2-\left({\rm ln} W_N)\right)_{y\tau},\\[3pt]
q[N]=\frac{1}{p[N]}\left((\frac{F_N}{W_N})_{\tau}-({\rm ln} W_N)_y({\rm ln} W_N)_{y\tau}\right),
\end{array}\right.
\end{equation}where $\nu_i,i=1,2,3$ are arbitrary constants and
\begin{equation}\label{phi}
\left\{\begin{array}{l}
\phi^1[N]=\frac{W(f_1,...,f_N,1)}{W_N},\\
\phi^2[N]=\frac{W(f_1,...,f_N,e^{\frac{1}{k}y})}{W_N},\\
\phi^3[N]=\frac{W(f_1,...,f_N,e^{-\frac{1}{k}y})}{W_N}.
 \end{array}\right.
\end{equation}
Direct calculations show that the asymptotic behaviours of the wave functions $\phi^1[N],\phi^2[N],\phi^3[N]$  are given by
\begin{eqnarray}\label{asy}
\phi^1[N]\sim\left\{\begin{array}{rl}
 (-1)^N\prod_{i=1}^{N}\alpha_i,\quad \quad \quad \quad \ (y\rightarrow +\infty),\\[8pt]
(-1)^N\prod_{i=1}^{N}\beta_i, \quad  \quad \quad \quad \ (y\rightarrow -\infty),\end{array} \right.\nonumber\\
\phi^2[N]\sim\left\{\begin{array}{rl}
e^{\frac{1}{k}y}\prod_{i=1}^{N}(\frac{1}{k}-\alpha_i), \quad \quad  \quad \ (y\rightarrow +\infty),\\[8pt]
e^{\frac{1}{k}y}\prod_{i=1}^{N}(\frac{1}{k}-\beta_i), \quad \quad  \quad \  (y\rightarrow -\infty),\end{array} \right.\\
\phi^3[N]\sim\left\{\begin{array}{rl}
e^{-\frac{1}{k}y}\prod_{i=1}^{N}(-\frac{1}{k}-\alpha_i),  \quad \quad (y\rightarrow +\infty),\\[8pt]
e^{-\frac{1}{k}y}\prod_{i=1}^{N}(-\frac{1}{k}-\beta_i),  \quad \quad  (y\rightarrow -\infty).\end{array} \right.\nonumber
\end{eqnarray}

Next, we work out the coordinate transformation between the independent variables  $x, t$ and $y, \tau $. To this end, we consider the spectral problem in (\ref{lax-1}) at
$\lambda=0$ which yields
\begin{equation}\label{lambda0}
\varphi_{xx}-\varphi=0, 
\end{equation}for $\varphi=\varphi_3$.
As $ {e^{x}}$ and ${e^{-x}}$ form a fundamental set of solutions of (\ref{lambda0}), in view of the gauge transformation (\ref{gu2}), they may be represented as
\begin{eqnarray*}
e^x&=\left(\frac{w}{h}\right)^{1/2}\left(c_1\phi^1[N]+c_2\phi^2[N]+c_3\phi^3[N]\right),\\
e^{-x}&=\left(\frac{w}{h}\right)^{1/2}\left(d_1\phi^1[N]+d_2\phi^2[N]+d_3\phi^3[N]\right),
\end{eqnarray*}
where the coefficients $c_k, d_k ~(k=1,2,3)$ are independent of $y$. Taking account of the asymptotic behaviours (\ref{asy}), we find
\begin{equation*}\label{spaceex}
 e^x=c_2 \left(\frac{w}{h}\right)^{1/2}\phi^2[N], \quad \quad
  e^{-x}=d_3\left(\frac{w}{h}\right)^{1/2}\phi^3[N],
\end{equation*}
which further imply
\begin{equation*}\label{spacex}
x= c+\frac{1}{2}{\rm ln}\left(\frac{\phi^2[N]}{\phi^3[N]}\right),
\end{equation*}
where $c=c(\tau)$. Differentiating above equation with respect to $\tau$ and
using the reciprocal transformation (\ref{rec1}), we have
\[
x_{\tau}=c_\tau+\frac{1}{2}\left(\ln\frac{\phi^2[N]}{\phi^3[N]}\right)_\tau
=uv,
\]
then by taking the limit $y\to \infty$, we find $c_\tau=u_0v_0=k^3$, which leads to $c=k^3\tau+d$ with $d$ as an integration constant.
Thus we obtain
\begin{equation}\label{x-y}
x=k^3\tau+\frac{1}{2}{\rm ln}\left(\frac{\phi^2[N]}{\phi^3[N]}\right)+d.
\end{equation}
For temporal variables, form (\ref{rec1}) we have
\begin{equation}\label{t-tau}
t=\tau.
\end{equation}
%

Finally, we need to work out the transformation formulae for the field variables $u, v$, which can be done by means of (\ref{p-q}) and the reciprocal transformation (\ref{rec1}). In particular,  we may deduce $w$ from $q[N]=\frac{w_y}{w}$. 
It is interesting to observe that for $p[N]$ and $q[N]$,  we have
\begin{align}\label{p1}
p[N]&=W(g_1,...,g_N)W(f'_1,...,f'_N)\left(\frac{k}{W_N}\right)^2,\\   \label{q1}
q[N]&=-\left(\ln\frac{W(g_1,g_2, ..., g_N)}{W_N}\right)_y,
\end{align}where $g_i=\lambda_i^{-2}(f_{iyy}-k^{-2}f_i), i\geq 1$. The proof of (\ref{p1}) and (\ref{q1}) is presented in the appendix.
Summarizing above discussions, we have
\begin{prop}
 The Geng-Xue equation admits the parametric representation of the multi-kink solutions
\begin{eqnarray}\label{Nsoliton}
u=k^2\ell W(f'_1,...,f'_N)\left(\frac{x_y}{W_NW(g_1,g_2, ..., g_N)} \right)^{\frac{1}{2}},\quad v=\frac{x_{\tau}}{u},\nonumber
\end{eqnarray}where $x, t$ are defiend by (\ref{x-y}) and (\ref{t-tau}), $\ell$ is an integration constant, and $\phi^2[N],\phi^3[N]$ are given by  (\ref{phi}),  respectively.
\end{prop}
\bigskip
%

In the rest part of this section, we consider the simplest cases and present two examples.

\noindent{\bf Example 1: 1-kink solution}

For $N=1$, let us take
\begin{equation*}
 f_{1}=e^{\vartheta_1}+\delta_1e^{\sigma_1},\quad    g_1=\lambda_1^{-2}(f_{1yy}-k^{-2}f_1),
\end{equation*}where $\delta_1$ may be chosen as $\pm 1$. Direct computations show that
the DT (\ref{dt1}) yields
\begin{align*}
\phi^2[1]&=\left(\frac{1}{k}-\frac{f_{1y}}{f_1}\right)e^{\frac{1}{k}y},\quad  \phi^3[1]=-\left(\frac{1}{k}+\frac{f_{1y}}{f_1}\right)e^{-\frac{1}{k}y},\\
p[1]&=k^2-k^2\left(\frac{g_1}{f_1}\right)_y,\quad \;\;    q[1]=-\frac{f_1}{g_1}\left(\frac{g_1}{f_1}\right)_y.
\end{align*}
Then, application of Proposition 2 allows us to have the following  solution of the Geng-Xue equation
\begin{align}\label{1soliton}
x&=\frac{1}{k}y+k^3\tau+\frac{1}{2}{\rm ln}\left(\bar{k}_1\frac{f_{1y}-\frac{1}{k}f_1}{f_{1y}+\frac{1}{k}f_1}\right),\quad t=\tau,\nonumber\\
u&=k^2\frac{f_{1y}}{f_1}\left(\bar{k}_2\frac{f_1}{g_1}x_y\right)^{\frac{1}{2}},\\
 v&=\frac{k^3f_{1y}\left(f_{1y}-\frac{1}{k^2}g_1\right)}{\left(f_{1y}-\frac{1}{k}f_1\right)\left(f_{1y}+\frac{1}{k}f_1\right)u}.\nonumber
\end{align}Here and in the sequel, all $\bar{k}_i,i\in \mathbb{Z}$ are assumed to be arbitrary constants. We take $\alpha_1,\beta_1,\lambda_1^2$ as  real constants so that we have the  real-valued solutions. Furthermore, assuming that $\delta_1=1, 0<k p_1<1$,  our solutions will be non singular.
Under these assumptions and  substituting the expressions of $f_1,g_1$ into the solution (\ref{1soliton}), we obtain a parameter representation of 1-kink solution of the Geng-Xue equation
\begin{align}
  x&=\bar{k}_3+\frac{1}{k}y+k^3\tau+\frac{1}{2}{\rm ln}\frac{\left(\frac{2}{k}-r_1\right){\rm cosh}\theta_1-p_1{\rm sinh}\theta_1}{\left(\frac{2}{k}+r_1\right){\rm cosh}\theta_1+p_1{\rm sinh}\theta_1},
\quad  t=\tau,\nonumber\\
 u&=\frac{\bar{k}_4(p_1{\rm sinh}\theta_1+r_1{\rm cosh}\theta_1)}{\sqrt{r_1^2{\rm cosh}(2\theta_1)-p_1r_1{\rm sinh}(2\theta_1)+p_1^2+r_1^2}},\\
 v&=\frac{k^3r_1[3k^2p_1r_1{\rm sinh}\theta_1-(k^2p_1^2+2){\rm cosh}\theta_1]}{\bar{k}_4(k^2p_1^2-1)\sqrt{r_1^2{\rm cosh}(2\theta_1)-p_1r_1{\rm sinh}(2\theta_1)+p_1^2+r_1^2}},\nonumber
\end{align}where $\bar{k}_3={\rm ln}(- \bar{k})$ and $\theta_1=\frac{1}{2}(p_1y+\frac{3k^2p_1}{p_1^2-k^{-2}}\tau+\theta_{10}),\ r_1=\alpha_1+\beta_1=\pm\sqrt{\frac{4}{3k^2}-\frac{1}{3}p_1^2}$
with $\theta_{10}$ any constant.
\begin{figure}[htbp]
\centering
\begin{minipage}[t]{0.48\textwidth}
\centering
\includegraphics[height=5cm,width=7cm]{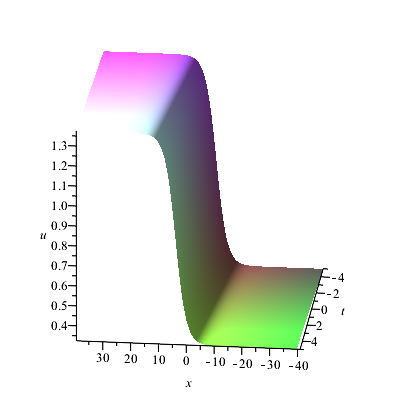}
\caption{One-kink for $u$ at $\bar{k}_3=\theta_{10}=0,p_1=1,k_1=\frac{1}{2},\bar{k}_4=1,r_1=\sqrt{5}$.}
\end{minipage}\quad
\begin{minipage}[t]{0.48\textwidth}
\includegraphics[height=5cm,width=7cm]{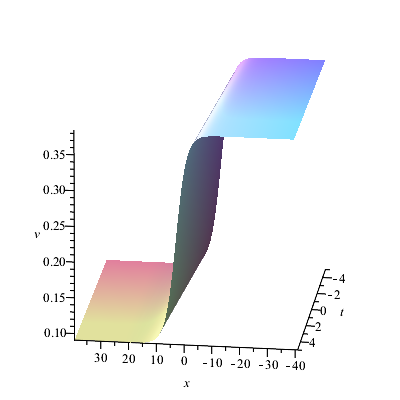}
\caption{One-antikink for $v$ at $\bar{k}_3=\theta_{10}=0,p_1=1,k_1=\frac{1}{2},\bar{k}_4=1,r_1=\sqrt{5}$. }
\end{minipage}
\end{figure}

\bigskip

\noindent{\bf Example 2: 2-kink}

For $N=2$, we may take
\begin{equation*}
f_{i}=e^{\vartheta_i}+\delta_ie^{\sigma_i},\quad   g_i=\frac{1}{\lambda_i^2}(f_{iyy}-k^{-2}f_i),\quad  \quad \quad i=1,2,
\end{equation*}where $\delta_1,\delta_2$ are allowed to be $\pm1$. From (\ref{dt2}), it follows that
\begin{align*}
\phi^2[2]&=\frac{W(f_1,f_2,e^{\frac{1}{k}y})}{W(f_1,f_2)}, \quad\quad   \phi^3[2]=\frac{W(f_1,f_2,e^{-\frac{1}{k}y})}{W(f_1,f_2)},\\
p[2]&=k^2-[{\rm ln} W(f_1,f_2)]_{y\tau}, \ q[2]=\frac{1}{p[2]}\left(\frac{W(f_{1y},f_{2y})}{W(f_1,f_2)}-\frac{1}{2}\left(\frac{W_y(f_1,f_2)}{W(f_1,f_2)}\right)^2\right)_{\tau}.
\end{align*}Noticte that $g_{iy}=f_i,f_{i\tau}=k^2 g_i, i=1,2$. Then, after some direct calculations, we find
\begin{align*}
\frac{\phi^2[2]}{\phi^3[2]}&=e^{\frac{2}{k}y}\frac{G_1}{G_2},\\
p[2]&=k^2\frac{W(g_1,g_2)W(f_{1y},f_{2y})}{W^2(f_1,f_2)},  \quad
q[2]=\left({\rm ln}\frac{W(f_1,f_2)}{W(g_1,g_2)}\right)_y,
\end{align*}where
\begin{eqnarray*}
G_1=W(f_{1y}-\frac{1}{k}f_1,f_{2y}-\frac{1}{k}f_2),\quad   G_2=G_1|_{k\rightarrow -k}.
\end{eqnarray*}Following the Proposition 2 and after tedious calculations,  we may get a parameter representation of exact solution
\begin{align}
x&=\bar{k}_5+\frac{1}{k}y+k^3\tau+\frac{1}{2}{\rm ln}\frac{G_1}{G_2},\quad t=\tau,\nonumber\\
u&=\bar{k}_6\frac{W(f_{1y},f_{2y})}{\sqrt{G_1G_2}},\\
v&=\frac{k^3}{\bar{k}_6\sqrt{G_1G_2}}W(f_{1y}-\frac{1}{k^2}g_1,f_{2y}-\frac{1}{k^2}g_2).\nonumber
\end{align}
 Hereafter, to obtain real solution without singularity, let us assume that $\alpha_i,\beta_i,\lambda_i^2$ are real and $\delta_1=\delta_2=1$.
Then we may establish the parameter representation of the 2-kink solution of the Geng-Xue equation. A profile of 2-kink solution is plotted in Fig. 3,4 for the parameter $p_1=2,p_2=\frac{3}{2},r_1=\alpha_1+\beta_1=\sqrt{7},r_2=\alpha_2+\beta_2=-\sqrt{\frac{91}{12}},\bar{k}_5=\theta_{10}=\theta_{20}=0,k=\frac{2}{5},\bar{k}_6=1$.

\begin{figure}[htbp]
\centering
\begin{minipage}[t]{0.48\textwidth}
\centering
\includegraphics[width=7cm]{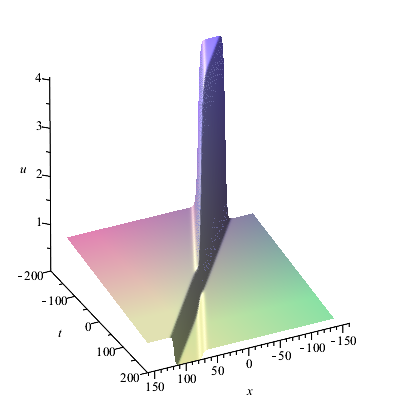}
\caption{two-kink  for $u$ .}
\end{minipage}\quad
\begin{minipage}[t]{0.48\textwidth}
\includegraphics[width=7cm]{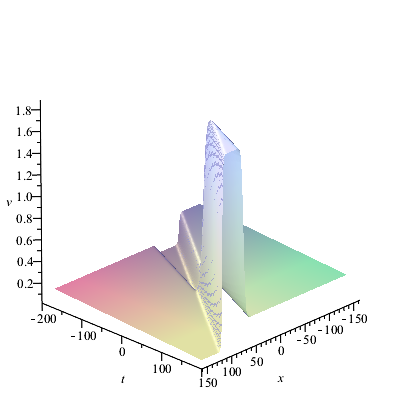}
\caption{two-kink  for $v$. }
\end{minipage}
\end{figure}
\begin{figure}[htbp]
\centering
\begin{minipage}[t]{0.48\textwidth}
\centering
\includegraphics[width=8cm]{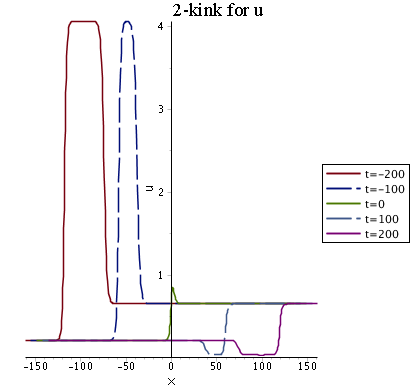}
\caption{two-kink solution for $u$ at $\tau=-200, -100, 0, 100, 200$.}
\end{minipage}\quad \
\begin{minipage}[t]{0.48\textwidth}
\includegraphics[width=8cm]{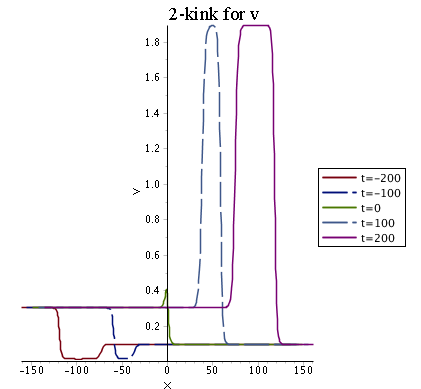}
\caption{two-kink solution for $v$ at $\tau=-200, -100, 0, 100, 200$. }
\end{minipage}
\end{figure}
%

\bigskip
\noindent
\section*{Acknowledgements}
This work is partially supported by the National Natural Science Foundation of China (Grant Nos. 11805071, 11871471 and 11931107).
 N. L. is grateful to HSE for supporting his visit  during Aug. 2019 to Sep. 2020, and especially to thank Ian Marshall and Maxim Pavlov for their hospitality.

 \appendix
 \setcounter{equation}{0}
\renewcommand\theequation{A.\arabic{equation}}
\section*{{Appendix A. \bf{Proof of (\ref{p1}) and (\ref{q1})}}}
For $N=1, 2$, the validity of (\ref{p1}, \ref{q1}) has been shown in the example 1 and example 2. For convenice, we introduce the following notations
\begin{equation*}
M=\left(
    \begin{array}{ccc}
     f'_1 & \cdots  & f_1^{(N-2)} \\
      \vdots & \vdots& \vdots \\
      f'_N & \cdots  & f_N^{(N-2)} \\
    \end{array}
  \right),\quad  \bar{M}=\left(
    \begin{array}{cccc}
     f'_1 & \cdots  & f_1^{(N-3)} & f_1^{(N-1)} \\
      \vdots & \vdots& \vdots & \vdots \\
      f'_N & \cdots  & f_N^{(N-3)} & f_N^{(N-1)} \\
    \end{array}
  \right),
\end{equation*}
and
\begin{equation*}
(\mathbf{g},\mathbf{f},\mathbf{a}, \mathbf{b},\mathbf{d})=\left(
                            \begin{array}{ccccc}
                              g_1  & f_1 & f_1^{(N-2)}& f_1^{(N-1)} &  f_1^{(N)}  \\
                              \vdots & \vdots & \vdots&\vdots &\vdots \\
                               g_N  & f_N & f_N^{(N-2)}& f_N^{(N-1)} &  f_N^{(N)}  \\
                            \end{array}
                         \right).
\end{equation*}

We assume $N\geq3$ and note $f_{j\tau}=k^2g_j$, then we deduce
\begin{align*}
p[N]&=k^2-k^2\left(\frac{\left|\mathbf{g},M,\mathbf{b}
                   \right|}{W_N}\right)_y\\
 &=\frac{k^2}{W_N^{2}}\big(\left| \mathbf{g},M,\mathbf{b}\right|W_{Ny}-\left|\mathbf{g},M,\mathbf{d}\right|W_{N}\big)\\
&=\frac{k^2}{W_N^{2}}\big(\left|M,\mathbf{g},\mathbf{b}\right|\left|M,\mathbf{f},\mathbf{d}\right|-\left|M,\mathbf{f},\mathbf{b}\right|\left|M,\mathbf{g},\mathbf{d}\right|)\\
&=\frac{k^2}{W_N^{2}}\left|M,\mathbf{g},\mathbf{f} \right|\left|M,\mathbf{b},\mathbf{d} \right|\\
&=\frac{k^2}{W_N^{2}}W(g_1,...,g_N)W(f'_1,...,f'_N),
\end{align*}
where the identity
\begin{equation}\label{idd}
\left|M,\mathbf{g},\mathbf{f}\right|\left|M,\mathbf{b},\mathbf{d}\right|-\left|M,\mathbf{g},\mathbf{b}\right|
\left|M,\mathbf{f},\mathbf{d}\right|+\left|M,\mathbf{g},\mathbf{d}\right|\left|M,\mathbf{f},\mathbf{b}\right|=0
\end{equation}
is used.

Next we consider (\ref{q1}). Indeed, we have
\begin{eqnarray*}
&&q[N]=\frac{1}{p[N]}\big((F_N/W_N)_{\tau}-({\rm ln} W_N)_y({\rm ln} W_N)_{y\tau}\big)\\
&&\hspace{0.8cm}=\frac{1}{p[N]}\left(\frac{W_{Ny}(p[N]-k^2)}{W_N}+\frac{F_{N\tau}W_N-F_NW_{N\tau}}{W_N^{2}}\right)\\
&&\hspace{0.8cm}=\frac{W_{Ny}}{W_N}+\frac{F_{N\tau}W_N-F_NW_{N\tau}-k^2W_NW_{Ny}}{k^2\hat{W}_N}\\
&&\hspace{0.8cm}=\frac{W_{Ny}}{W_N}+\frac{\left|\bar{M},\mathbf{g},\mathbf{d}\right|\left|\bar{M},\mathbf{a},\mathbf{f}\right|
+\left|\bar{M},\mathbf{g},\mathbf{a}\right|\left|\bar{M},\mathbf{f},\mathbf{d}\right|}{\hat{W}_N}\\
&&\hspace{0.8cm}=\frac{W_{Ny}}{W_N}+\frac{\left|\bar{M},\mathbf{a},\mathbf{d}\right|\left|\bar{M},\mathbf{g},\mathbf{f}\right|}{\hat{W}_N}\\
&&\hspace{0.8cm}=\frac{W_{Ny}}{W_N}-\frac{W_y(g_1,...,g_N)}{W(g_1,...,g_N)},
\end{eqnarray*}
where $\hat{W}_N=W(g_1,...,g_N)W(f'_1,...,f'_N)$ and an identity similar to (\ref{idd}) is used again.


\bibliographystyle{sigma}
\bibliography{reference}

\end{document}